\documentclass[12pt]{iopart}

\usepackage{graphicx} 
\usepackage{dcolumn}
\usepackage{bm}
\usepackage{amssymb}
\usepackage{color}

\begin{document}
\bibliographystyle{unsrt}


\title[The Biorthogonal Decomposition and the identification of zonal flows]{The use of the Biorthogonal Decomposition for the identification of zonal flows at TJ-II}

\author{B.Ph.~van Milligen, E.~S\'anchez, A.~Alonso, M.A.~Pedrosa, C. Hidalgo, A.~Mart\'in de Aguilera, A.~L\'opez Fraguas}
 \address{Laboratorio Nacional de Fusi{\'o}n, CIEMAT, Avda.~Complutense 40, 28040 Madrid, Spain}

\date{\today}

\begin{abstract}
This work addresses the identification of zonal flows in fusion plasmas.
Zonal flows are large scale phenomena, hence multipoint measurements taken at remote locations are required for their identification.
Given such data, the Biorthogonal Decomposition (or Singular Value Decomposition) is capable of
extracting the globally correlated component of the multipoint fluctuations.
By using a novel quadrature technique based on the Hilbert transform, propagating global modes (such as MHD modes) can be distinguished from the non-propagating, synchronous (zonal flow-like) global component.
The combination of these techniques with further information such as the spectrogram and the spatial structure then allows an unambiguous identification of the zonal flow component of the fluctuations.
The technique is tested using gyro-kinetic simulations. 
The first unambiguous identification of a zonal flow at the TJ-II stellarator is presented, based on multipoint Langmuir probe measurements.
\end{abstract}

\maketitle

\section{Introduction}

Spontaneous confinement transitions occurring in magnetically confined plasmas are vital for the development of fusion as a viable future energy source.
It is generally assumed that Zonal Flows are an essential ingredient for understanding these transitions~\cite{Wagner:2007}.
These flows can be generated spontaneously from turbulence via Reynolds stress~\cite{Burrell:1997}.
However, the experimental identification of zonal flows is a hard problem.

Zonal flows are low-frequency phenomena with a global character, i.e., with long wavelengths in the toroidal and poloidal directions~\cite{Calvo:2009}.
More specifically, they are associated with a potential perturbation with toroidal mode number $n=0$,  poloidal mode number $m = 0$ and finite radial wavelength $k_r$.
The detection of these flows has been difficult, partly due to the fact that fusion-grade plasmas are characterized by the presence of many instabilities and waves, making it hard to isolate the low-frequency, long-wavelength zonal flow in the data, containing other fluctuating and/or oscillating contributions~\cite{Fujisawa:2007b,Fujisawa:2009}.
Several techniques have been {used}, including the measurement of spectra and (long-range) correlation
~\cite{McKee:2003,Pedrosa:2005,Pedrosa:2005c,Bencze:2006,Zhao:2006,Pedrosa:2007,Pedrosa:2008,Xu:2009,Liu:2009}, the Hilbert-Huang transform~\cite{Carreras:2011}, bicoherence~\cite{Diamond:2000,Tynan:2001,Milligen:2008c} and non-linear energy transfer~\cite{Xu:2012}.

The fact that zonal flows are `global' in the sense that they are long-wavelength phenomena should facilitate their unambiguous detection provided simultaneous measurements at several remote points inside the plasma are available, at the (radial) location where the zonal flow exists.

Given such multipoint data, a technique is needed to isolate and extract the `global' component from the fluctuation data.
This task is ideally suited for the BiOrthogonal Decomposition (BOD)~\cite{Dudok:1994},
also known as Proper Orthogonal Decomposition (POD) or Singular Value Decomposition (SVD), among others~\cite{Kerschen:2005}.
However, there may be other global oscillations affecting the fluctuation data, different from Zonal Flows.
Therefore, in the present work, the BOD technique is complemented with additional techniques to quantify the long-range character of the biorthogonal modes (cf.~\cite{Alonso:2012}) or their propagating nature, thus facilitating the distinction between global modes associated with Magneto-HydroDynamic (MHD) activity or other oscillations, on the one hand, and proper Zonal Flows, on the other. 

\section{The biorthogonal decomposition}\label{Method}

The details of the biorthogonal decomposition are well-known, and the reader is referred to the references for detailed information~\cite{Dudok:1994,Kerschen:2005}. Here, we will summarize its main features.

The multipoint measurements constitute a data matrix  $Y(i,j)$, where the index $i = 1,...,N$ labels the time and $j = 1,...,M$ the detector. 
Typically, the time corresponding to the time index, $t(i)$, is equally spaced, since measurements are typically taken at a fixed sampling rate, although this is not strictly necessary.
The physical location of the detectors, $\vec x(j)$, however, is often dictated by practical convenience or space limitations and will not necessarily correspond to a regular grid.
Finally, to facilitate the physical interpretation of the results, the measurements $Y$ performed at the various detectors $j$ should ideally be of the same physical quantity (in our case, an electric potential) and be given in the same units.

The BOD method decomposes the data matrix $Y(i,j)$ as follows:
\begin{equation}\label{biortho}
Y(i,j) = \sum_k \lambda_k \psi_k(i) \phi_k(j) 
\end{equation}
where $\psi_k$ is a `chrono' (a temporal function) and $\phi_k$ a `topo' (a spatial or detector-dependent function), such that the chronos and topos satisfy the following orthogonality relation~\cite{Dudok:1994}:
\begin{equation}\label{norm}
\sum_i{\psi_k(i)\psi_l(i)} = \sum_j{\phi_k(j)\phi_l(j)} = \delta_{kl}. 
\end{equation}
An alternative normalization (indicated by the superscript $r$) is 
\begin{equation}\label{norm_rms}
\frac{1}{N}\sum_i{\psi^r_k(i)\psi^r_l(i)} = \frac{1}{M}\sum_j{\phi^r_k(j)\phi^r_l(j)} = \delta_{kl}. 
\end{equation}
such that the root-mean-square (RMS) value of the topos and chronos equals 1.
In this case,
\begin{equation}\label{biortho_rms}
Y(i,j) = \sum_k \lambda^r_k \psi^r_k(i) \phi^r_k(j) 
\end{equation}
with $\lambda^r_k = \lambda_k/\sqrt{NM}$. 
With this normalization, the number $\lambda_k^r$ represents the contribution of mode $k$ to the total RMS `fluctuation amplitude'.
Thus, the fractional contribution of a given mode $k$ to the total `fluctuation energy' (proportional to the square of the RMS) can be computed from
\begin{equation}\label{mode_energy}
e_k^2 = \frac{\lambda_k^2}{ \sum_{k'} \lambda_{k'}^2 }
\end{equation}
The superscript `$r$' is irrelevant in this expression.
 
The combination chrono/topo at a given $k$, $\psi_k(i) \phi_k(j)$, is called a spatio-temporal `mode' of the fluctuating system, and is constructed from the data matrix without any prejudice regarding the mode shape.

The decomposition is performed by computing the Singular Value Decomposition (SVD) of the data matrix $Y(i,j)$. 
Note that this is always possible for any real-valued rectangular matrix $Y$, and standard software packages are available for this purpose.
The $\lambda_k \ge 0$ are the eigenvalues (sorted in decreasing order), where $k=1,...,\min(N,M)$.

A threshold may be set for cutting off the expansion in BOD modes (based on, e.g., a noise level) and so keeping only the dominant contribution of modes to the data matrix for further analysis and ignoring noisy or minor contributions.
When the BOD expansion is cut off at a certain mode $k_{\rm max}$, the reconstructed data, $Y_{\rm rec}$, are defined by Eq.~(\ref{biortho}) while restricting the sum to $k \le k_{\rm max}$.
In this case, the reconstructed data $Y_{\rm rec}$ minimize the least squares error with respect to the actual data $Y$ among all possible sets of $k_{\rm max}$ spatiotemporal modes. 

If the physical system under study contains normal mode oscillations and is sufficiently well-sampled (spatially and temporally), the probability that the BOD modes correspond closely to the said normal modes is high; the BOD is particularly sensitive to resonant linear normal modes~\cite{Kerschen:2005}.
On the other hand, the BOD technique is perhaps less suited for the analysis of non-linear systems in which no normal modes exist, as the BOD modes are unlikely to correspond to any meaningful physical modes in this case. 
An advantage of the BOD analysis is that no {\it a priori} assumption is made regarding the mode shape or spectral properties, unlike standard analysis techniques such as Fourier decomposition. 
This has the mentioned advantage that the BOD modes concentrate a maximum of fluctuation power in the lowest modes, but the disadvantage that the modes are not guaranteed to coincide exactly with the normal modes of the system (if known).

Furthermore, it should be noted that the SVD decomposition is not unique. First, the sign of the chrono and topo of a given mode $k$ can trivially be inverted without affecting their product, i.e., without affecting the reconstruction according to Eq.~(\ref{biortho}). 
Second, if two SVD modes $k_1$ and $k_2$ have the same eigenvalues, a rotation in the two-dimensional vector space spanned up by the chronos $k_1$ and $k_2$ and a compensating rotation in the space spanned up by the corresponding topos can be defined so that the reconstructed data, according to Eq.~(\ref{biortho}), are {unchanged}, giving rise to a (rotational) indeterminacy.
As will be clarified below, this may occur in the case of propagating modes. 
As such, it does not imply a great disadvantage, since such modes will be considered pairwise anyway.

\subsection{Covariance and the biorthogonal decomposition}

The covariance between a signal $Y(i,j_1)$ and a signal $Y(i,j_2)$ ($i$ being the temporal and $j$ the spatial index) is defined as:
\begin{equation}\label{covariance}
{\rm Cov}(j_1,j_2) = \frac{1}{N}\sum_i{Y(i,j_1)Y(i,j_2)}
\end{equation}
Since the signals are expanded according to Eq.~(\ref{biortho}), this expression can be simplified, using Eq.~(\ref{norm}), to:
\begin{eqnarray}\label{covariance_topos}
{\rm Cov}(j_1,j_2) &=& \frac{1}{N} \sum_k{\lambda_k^2\phi_k(j_1)\phi_k(j_2)} \nonumber \\
&=& \sum_k{(\lambda_k^r)^2\phi_k^r(j_1)\phi_k^r(j_2)},
\end{eqnarray}
i.e., the covariance between two signals is simply the sum of the products of the corresponding topos, weighed by the square of the eigenvalues. 
In other words, the topos reflect the covariance between the measurement signals; and the contribution of each mode $k$ to the covariance ${\rm Cov}(j_1,j_2)$ is simply $(\lambda_k^r)^2\phi_k^r(j_1)\phi_k^r(j_2)$.

From Eq.~(\ref{covariance_topos}) one immediately obtains, again using Eq.~(\ref{norm}):
\begin{equation}
\sum_{j_2}{{\rm Cov}(j_1,j_2)\phi_k(j_2)} = \frac{\lambda_k^2}{N}\phi_k(j_1).
\end{equation}
Thus, the topos $\phi_k$ are the eigenvectors of the covariance matrix ${\rm Cov}$, and $\lambda_k^2/N$ the corresponding eigenvalues.
This suggests that the topos $\phi_k$ and eigenvalues $\lambda_k$ can be found by computing the eigenvectors and eigenvalues of the covariance matrix ${\rm Cov}$, after which the corresponding chronos $\psi_k$ can be obtained by simple matrix multiplication~\cite{Kerschen:2005}:
\begin{equation}
\lambda_k\psi_k(i) = \sum_j{Y(i,j)\phi_k(j)}
\end{equation}
This view of the BOD mode decomposition may be helpful when interpreting the meaning of the BOD modes.

The fact that the topos are the eigenvectors of the covariance matrix immediately suggests an alternative approach: instead of analyzing the raw data $Y(i,j)$, one could first normalize the data to their RMS value
\begin{equation}
y(i,j) = \frac{Y(i,j)}{\sqrt{{\rm Cov}(j,j)}}
\end{equation}
When computing the SVD of $y(i,j)$ instead of $Y(i,j)$, the resulting topos $\phi_k$ will then be the eigenvectors of the {\it correlation} matrix $C$.

\subsection{Correlation}
By definition, the correlation between measurements $j_1$ and $j_2$ is obtained by normalizing the covariance, Eq.~(\ref{covariance}), to the RMS of each signal:
\begin{equation}
C(j_1,j_2) = \frac{{\rm Cov}(j_1,j_2)}{\sqrt{{\rm Cov}(j_1,j_1){\rm Cov}(j_2,j_2)}}
\end{equation}
From Eq.~(\ref{covariance_topos}), it is clear that this quantity only depends on the eigenvalues and the topos.
The contribution of a given mode $k$ to the correlation is (cf.~\cite{Alonso:2012})
\begin{equation}\label{correlation}
C_k(j_1,j_2) = \frac{(\lambda_k^r)^2\phi_k^r(j_1)\phi_k^r(j_2)}{\sqrt{{\rm Cov}(j_1,j_1){\rm Cov}(j_2,j_2)}}
\end{equation}
 
Thus, it is possible to quantify the contribution of a specific BOD mode $k$ to the so-called {\it Long Range Correlation} (LRC).
To achieve this, assume we may subdivide the set of measurements $j=1,\dots,M$ into two complementary sets: $\{j \in S_1\}$, with $M_1$ elements, and  $\{j \in S_2\}$, with $M_2$ elements, such that measurements in set $S_1$ are taken at a remote location from measurements in set $S_2$. 
The Long Range Correlation is then defined as the correlation between any pair of measurements taken at remote locations.
We define the contribution to the {\it mean} Long Range Correlation of a given mode $k$ by averaging the correlation due to this mode among any two remote measurement points:
\begin{equation}\label{LRC}
C_k^{\rm LR} = \frac{1}{M_1M_2}\sum_{\{j_1 \in S_1\}}\sum_{\{j_2 \in S_2\}}{C_k(j_1,j_2)}
\end{equation}
This definition may lead to low values in the case of partial cancellations due to the existence of a mixture of correlations and anti-correlations between individual measurements.
To measure the overall correlation `intensity' regardless of sign, we also define the mean absolute Long Range Correlation:
\begin{equation}\label{LRCabs}
C_k^{\rm LR, abs} = \frac{1}{M_1M_2}\sum_{\{j_1 \in S_1\}}\sum_{\{j_2 \in S_2\}}{\left | C_k(j_1,j_2) \right |}
\end{equation}
 

%

\subsection{Mode identification}

One important aspect of the BOD method is that it assumes that all 'modes' are separable in space and time. 
Standing waves are of this type.
However, one often has to contend with propagating structures, such as running waves.
It is important to be able to recognize and distinguish these different structures.

A typical example of a running wave is $Y(t,x) = \cos(kx-\omega t)$. 
This can trivially be decomposed into a sum of two standing wave patterns:
\begin{equation}
\cos(kx-\omega t) = \cos(kx)\cos(\omega t)-\sin(kx)\sin(\omega t)
\end{equation}
By Fourier's Theorem, this result can be generalized to any linearly propagating spatiotemporal structure of argument $kx-\omega t$.
Thus, linearly propagating or `traveling' waves or structures generate a pair of BOD modes, with similar eigenvalues and a mutual phase difference (in space and time) of around 90$^\circ$~\cite{Dudok:1994}.

Consequently, a method is needed to identify such mode pairs.
If the propagating mode has a clearly defined frequency $\omega$, the Fourier power spectra $|\hat \psi_k|^2$ of the concerned chronos will have a peak at this frequency (here, the circumflex indicates the Fourier transform), and the cross phase between the chronos $k_1,k_2$ is easily determined as the phase of the complex cross spectrum  $\hat \psi_{k_1} \hat \psi_{k_2}^*$ at the frequency $\omega$, where the star signifies complex conjugate.

As the low-frequency modes of interest do not always have a clearly defined spectral peak, we will use a technique for quadrature detection based on the Hilbert transform~\cite{Hahn:1996}.
In this framework, we must assume that the BOD topos and chronos fluctuate symmetrically about zero, which can be achieved by subtracting their mean or applying a suitable high-pass frequency filter prior to analysis.
This is necessary as it is a well-known fact that the Hilbert transform does not produce a reasonable estimate of the quadrature if the mean of the signal is not zero or is drifting~\cite{Hahn:1996}.
With this additional assumption, it is feasible to compute the quadrature of a given topo or chrono by means of the Hilbert transform.

Thus, we can use the full set of topos and chronos and compute the following spatial and temporal quadrature matrices:
\begin{eqnarray}
Q_x(k_1,k_2) = \sum_j \tilde \phi_{k_1}(j) H(\tilde \phi_{k_2}(j))\nonumber \\
Q_t(k_1,k_2) = \sum_i  \tilde \psi_{k_1}(i) H(\tilde \psi_{k_2}(i)) 
\end{eqnarray}
where $H$ is the Hilbert transform and the tilde refers to the removal of the mean (or lowest frequencies) mentioned above.
The elements of these quadrature matrices are restricted to the range $[-1,1]$; the absolute value of a given element $Q_{x,t}(k_1,k_2)$ will differ significantly from 0 when the corresponding modes are in approximate quadrature.
By means of this technique, it becomes possible to identify linearly propagating structures by finding pairs of modes with similar eigenvalues, $\lambda_{k_1} \simeq \lambda_{k_2}$, such that $Q_x(k_1,k_2)$ and $Q_t(k_1,k_2)$ are significantly different from zero.
By default, modes that do not occur in pairs then correspond to standing wave structures.

\clearpage
\section{Tests using gyrokinetic simulations}\label{Euterpe}

To test whether the analysis methods described in the preceding section are capable of delivering the promised results, in this section we will apply these techniques to gyrokinetic simulations with known properties, carried out with the code EUTERPE. 

EUTERPE is a global gyrokinetic particle-in-cell Monte Carlo code developed originally at CRPP Lausanne~\cite{Jost:2001} with the aim of simulating plasma turbulence in arbitrary threedimensional geometries.  It is being developed at the Max Plank IPP (Greifswald, Germany)~\cite{Kornilov:2005,Kauffmann:2010,Kleiber:2011,Kleiber:2012} and is presently used at the National Fusion Laboratory (CIEMAT, Spain) for the simulation of TJ-II plasmas. For more {information} about the code the reader is referred to the cited references.

Several linear simulations have been run in TJ-II geometry: linear simulations of zonal flow relaxation~\cite{Sanchez:2013} and simulations of Ion Temperature Gradient (ITG) instabilities~\cite{Sanchez:2011}. Results from these two kinds of simulations are used in this work to test the zonal flow detection technique.

\begin{figure} \centering
  \includegraphics[trim=0 100 0 0,clip=,width=16cm]{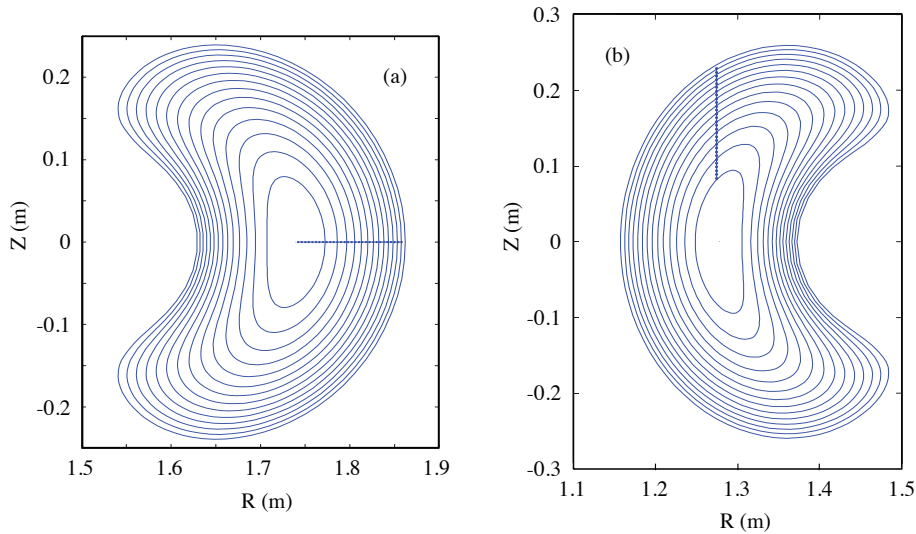}
\caption{\label{Euterpe_probe}Euterpe synthetic probes. (a): $\phi = 0^\circ$. (b): $\phi = 45^\circ$. In both poloidal cross sections, 30 equidistant synthetic probes (dots) pick up the electric plasma potential.}
\end{figure}

\subsection{Simulation of Zonal Flow relaxation}

In this kind of simulation, the relaxation of an initial zonal perturbation to the density is studied following the seminal work by Rosenbluth and Hinton~\cite{Rosenbluth:1998}. The simulation is initiated with a perturbation to the density of the form $\delta n_i \propto n_0 \langle k_r^2 \rho_i^2/T_i \rangle \cos (4 \pi s)$, where $s$ is the normalized toroidal flux used as radial coordinate in EUTERPE, $n_0$ the equilibrium density, $k_r= 4 \pi |\nabla s |$ the radial wavenumber of the perturbation, $\rho_i=\sqrt{2m_i T_i }/eB$ the ion Larmor radius, $m_i$ and $e$ the ion mass and charge, respectively, $T_i$ the equilibrium ion temperature, $B$ the magnetic field and $\langle \rangle$ means flux surface average.
The simulation is carried out in the standard magnetic configuration of TJ-II (labelled 100\_44\_64) under plasma conditions in which a neoclassical root confinement transition (the so-called low density transition) occurs at the plasma edge.
The simulation corresponds to a numeric experiment in which this root transition is simulated by slowly  evolving density and temperature profiles~\cite{Velasco:2013}. The density and temperature profiles and also the neoclassical equilibrium electric field are included in the simulation.

The initial perturbation is allowed to evolve linearly without taking into account collisions and the plasma potential is monitored at several positions emulating several multi-pin Langmuir probes measuring plasma potential at a set of radial positions, as shown in the Fig.~\ref{Euterpe_probe}: probe 1 is located at $\phi=0^\circ$ and probe 2 at half the period ($45^\circ$). The probe pins (30 for each probe) are ordered from large to small radius and cover most of the minor radius.

\begin{figure} \centering
  \includegraphics[trim=0 0 0 0,clip=,width=10cm]{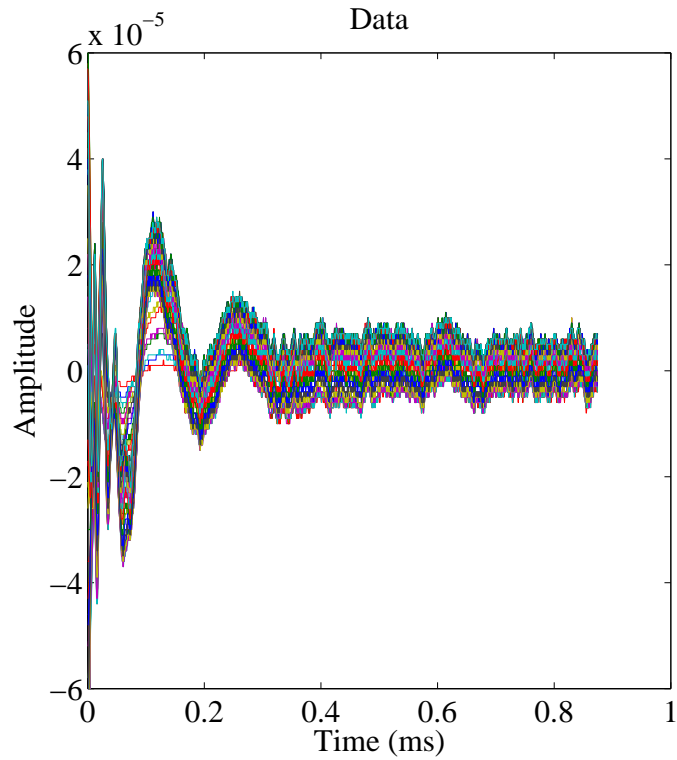}
\caption{\label{Euterpe_ZF_data}Raw data of an Euterpe calculation of Zonal Flow relaxation (60 probes).}
\end{figure}

In this simulation, the zonal component of the potential ($m=0, n=0$) is observed to relax via a low frequency damped  oscillation, typical for stellarators~\cite{Mishchenko:2008}, and also higher frequency oscillations identified as Geodesic Acoustic Modes (GAM) appear (see \cite{Sanchez:2013} and references therein). The frequency of the slow oscillation is in the range of 5--10 kHz and the GAM, only noticeable at the very beginning of the simulation, is around $50-100$ kHz. 
The GAM oscillation is associated with low-$m$ modes having much lower amplitude than the $m=0$ component. 
Raw simulation data are shown in Fig.~\ref{Euterpe_ZF_data}.

\begin{figure} \centering
  \includegraphics[trim=0 0 0 0,clip=,width=10cm]{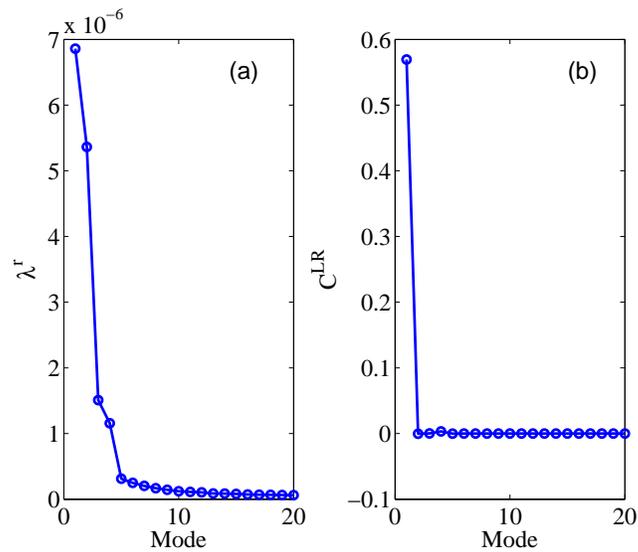}
\caption{\label{Euterpe_ZF_LRC}(a) BOD eigenvalues and (b) the long range correlation contribution of each mode, $C^{\rm LR}$, Euterpe calculation of Zonal Flow relaxation.}
\end{figure}

The BOD eigenvalues and the long range correlation contribution of each mode, $C^{\rm LR}$, are shown in Fig.~\ref{Euterpe_ZF_LRC}.
Clearly, nearly all LRC is concentrated in the first BOD mode.

\begin{figure} \centering
  \includegraphics[trim=50 0 50 0,clip=,width=16cm]{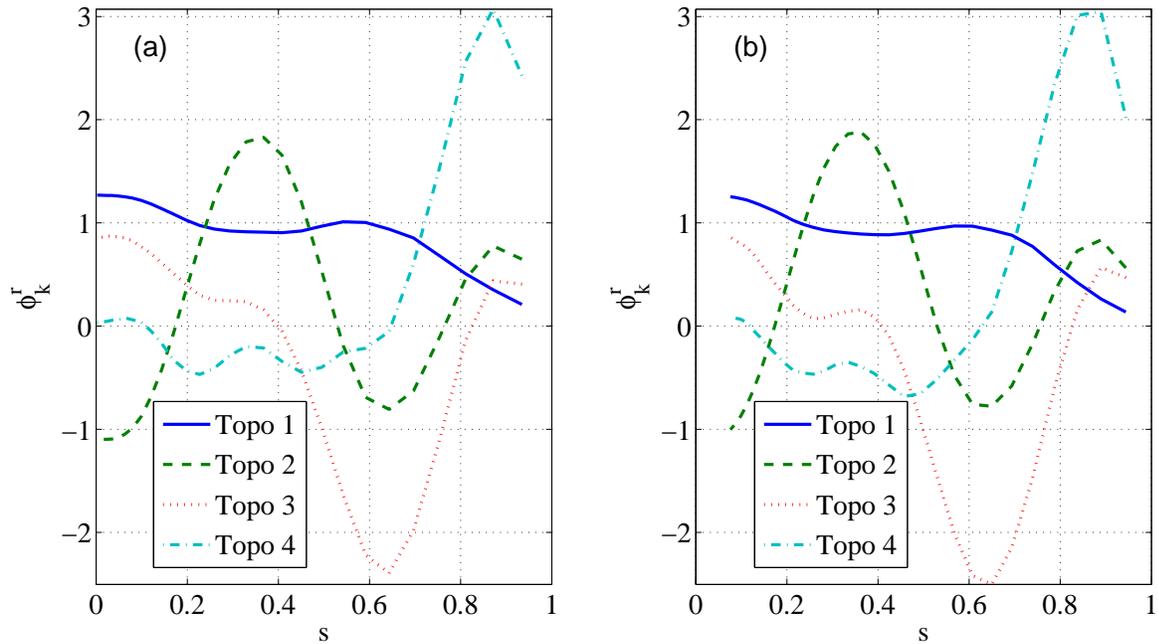}
\caption{\label{Euterpe_ZF_topos}BOD topos for (a) probe 1 and (b) probe 2, Euterpe calculation of Zonal Flow relaxation.}
\end{figure}

\begin{figure} \centering
  \includegraphics[trim=0 0 0 0,clip=,width=10cm]{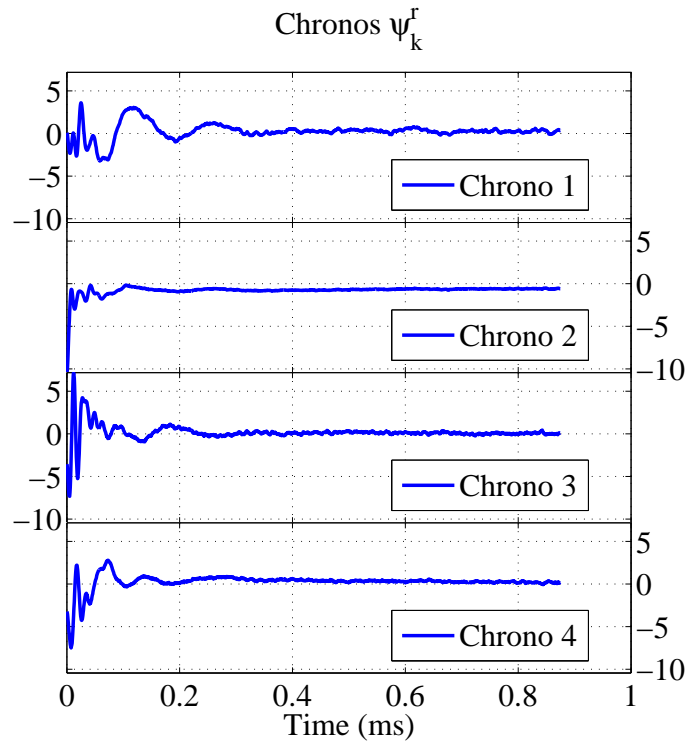}
\caption{\label{Euterpe_ZF_chronos}BOD chronos, Euterpe calculation of Zonal Flow relaxation.}
\end{figure}

\begin{figure} \centering
  \includegraphics[trim=0 0 0 0,clip=,width=7.5cm]{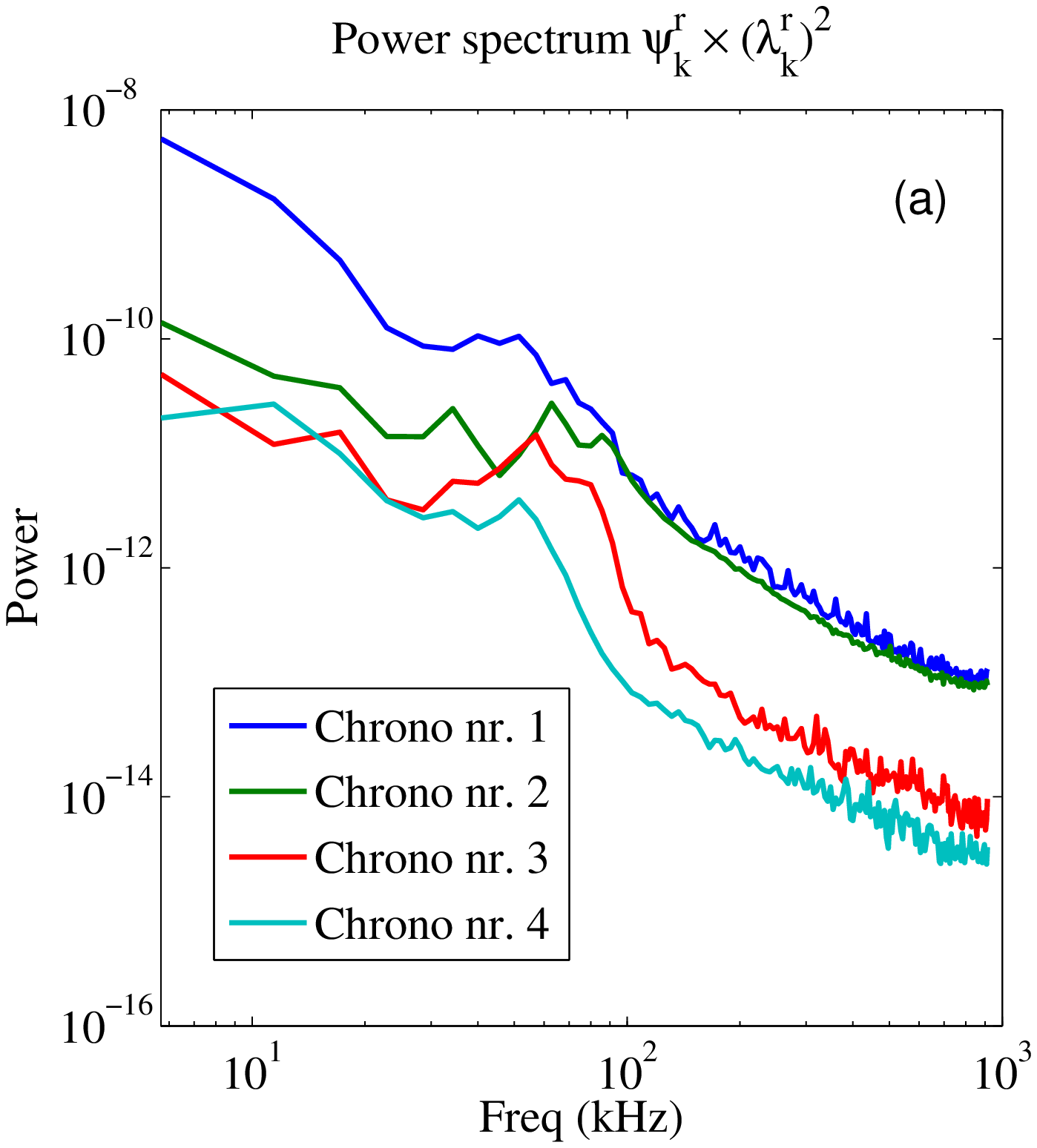}
  \includegraphics[trim=0 0 0 0,clip=,width=7.5cm]{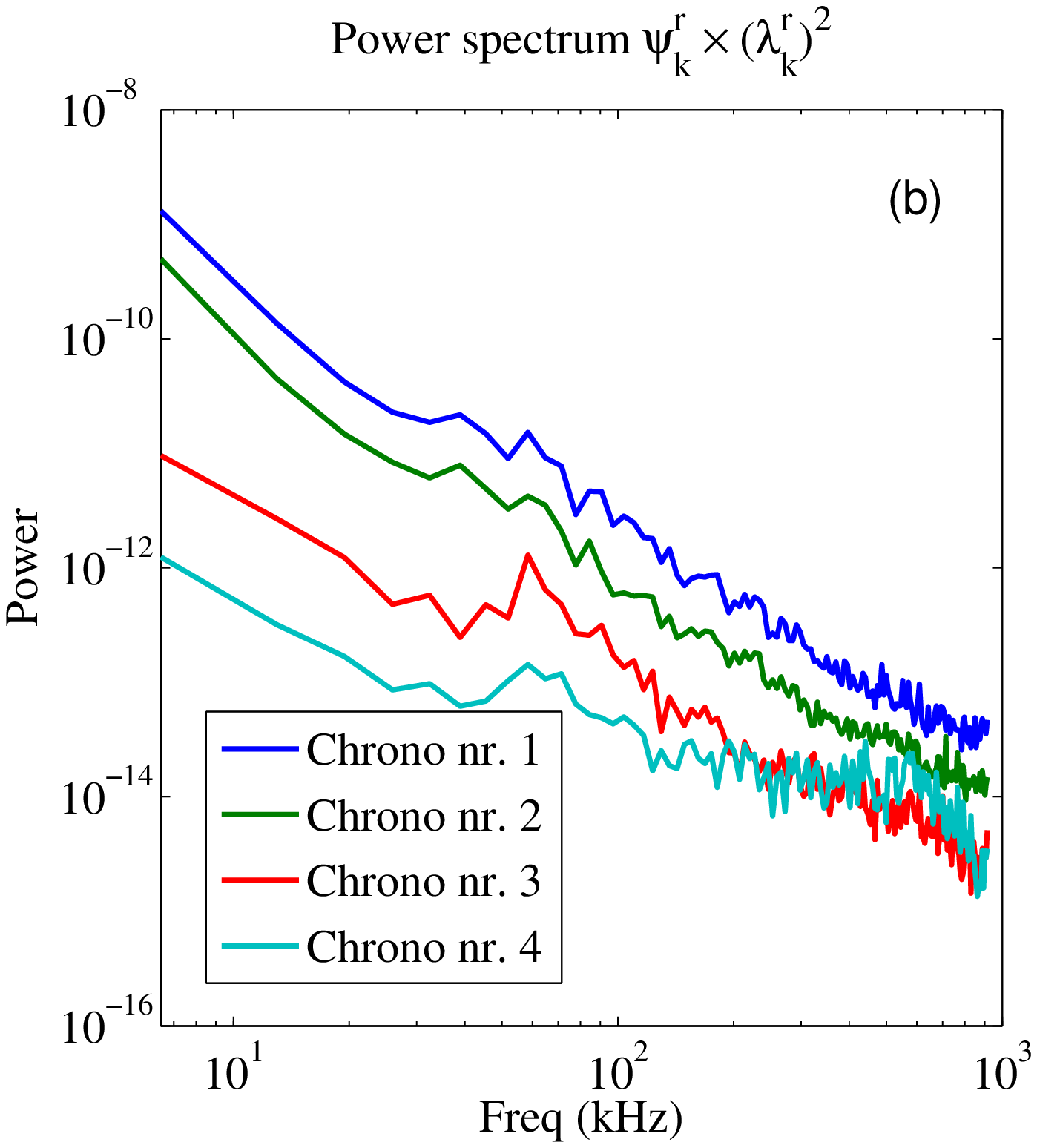}
\caption{\label{Euterpe_ZF_spectra}BOD chronos spectra, Euterpe calculation of Zonal Flow relaxation. 
(a): spectrum of the chronos over the whole time window; (b): spectrum of the chronos for $t\ge 0.1$ ms.}
\end{figure}

\begin{figure} \centering
  \includegraphics[trim=100 0 100 0,clip=,width=14cm]{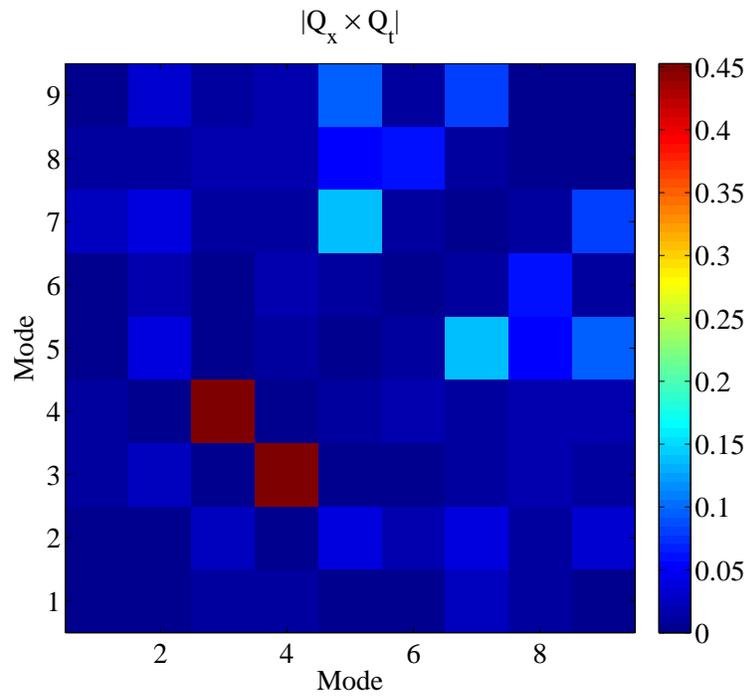}
\caption{\label{Euterpe_ZF_quadrature}Quadrature matrix of the first few BOD modes, Euterpe calculation of Zonal Flow relaxation.}
\end{figure}

Fig.~\ref{Euterpe_ZF_topos} shows the first four topos, Fig.~\ref{Euterpe_ZF_chronos} the first four chronos, and Fig.~\ref{Euterpe_ZF_spectra} the spectra of the first four chronos. Note that the spectrum of the first chrono exhibits the expected temporal behavior of a Zonal Flow: it peaks at very low frequency.
Comparing the spectrum calculated for the whole time window and for $t\ge 0.1$ ms, it is clear that the
peak at about 50 kHz corresponds mainly to the GAM appearing in the initial part of the time window ($t<0.1$ ms).
In this respect, the quadrature analysis, shown in Fig.~\ref{Euterpe_ZF_quadrature}, shows that modes 3 and 4, most clearly showing the mentioned spectral peak (Fig.~\ref{Euterpe_ZF_spectra}), form a propagating mode pair, as one might have suspected on the basis of their similar eigenvalues (Fig.~\ref{Euterpe_ZF_LRC}) and spectra. 
{This propagation would be {\it radial}, not poloidal/toroidal, in view of the near identity of the topos 3 and 4 in the two remote probes, which is consistent with the GAM nature of the fluctuations (constancy of potential on flux surfaces).}

\begin{figure} \centering
  \includegraphics[trim=0 0 0 0,clip=,width=16cm]{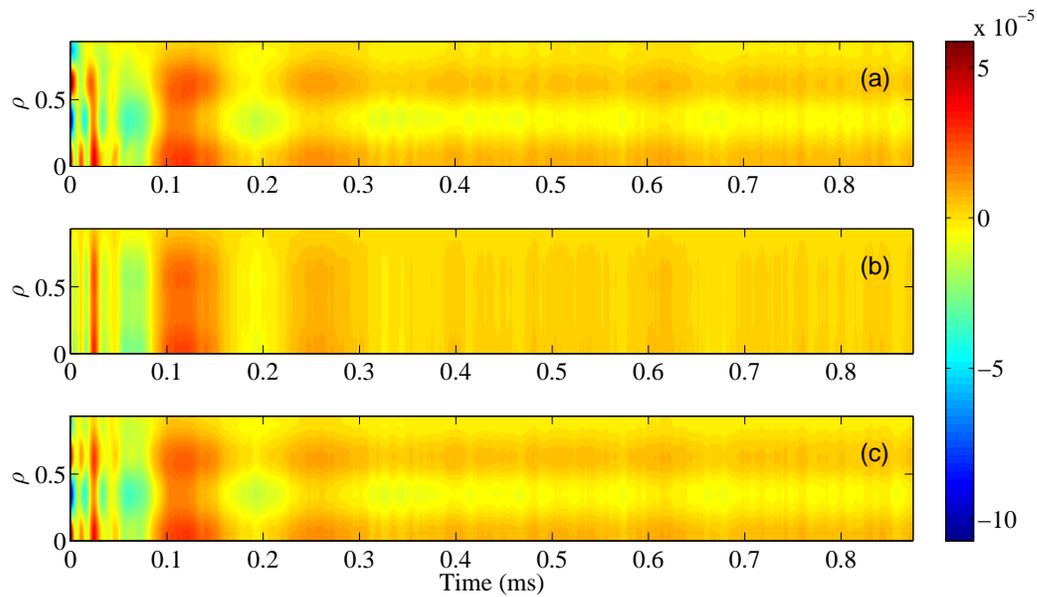}
\caption{\label{reconstruction}{(a) $\phi_{00}$, the flux surface averaged component of the simulated potential; (b) reconstructed potential oscillations using the first BOD mode; (c) reconstructed potential oscillations using the first two BOD modes.}}
\end{figure}

Thus having identified modes 1 and modes 3+4, mode 2 remains. 
Note that this radially oscillating mode with very similar shape in both probes (Fig.~\ref{Euterpe_ZF_topos}) hardly oscillates in time but mainly decays in amplitude (Fig.~\ref{Euterpe_ZF_chronos}), while exhibiting a very similar spectrum as mode 1 for $t > 0.1$ ms (Fig.~\ref{Euterpe_ZF_spectra}).
{We conclude that in this case, the Zonal Flow is captured by the first 2 BOD modes, with interesting properties: mode 1 has no radial sign changes and oscillates in time, while mode 2 has no temporal sign changes and oscillates in space.
Note that in view of the orthogonality requirements of the BOD modes, Eq.~(\ref{norm}), there can be only one topo that does not change sign and only one chrono that does not change sign.

Fig.~\ref{reconstruction} shows a comparison between the $\phi_{00}$, the flux surface averaged component of the simulated potential, and the reconstructed potential oscillations using only the first and the first two BOD modes, respectively. Very close agreement is observed when both ZF BOD modes are included.
}

\clearpage
\subsection{Linear simulation of an ITG instability}

In this case, a linear simulation of ideal ITG instability is studied in the TJ-II standard configuration~\cite{Sanchez:2011,Mishchenko:2008}.

\begin{figure} \centering
  \includegraphics[trim=0 0 0 0,clip=,width=10cm]{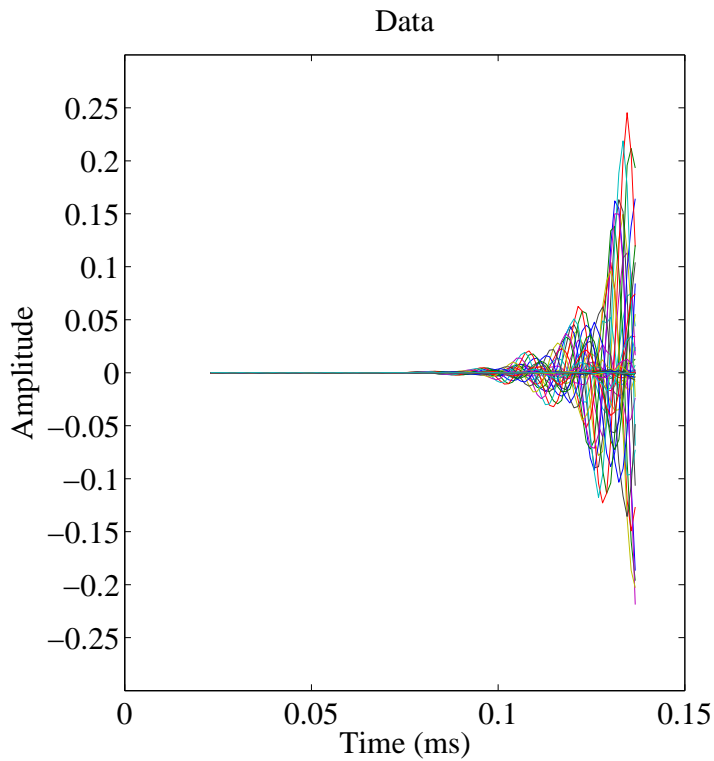}
\caption{\label{Euterpe_ITG_data}Raw data from an Euterpe simulation of ITG turbulence (60 probes).}
\end{figure}

The ion and electron density profiles and the electron temperature profiles are { taken to be} flat while the ion temperature profile { is defined to have a typical tanh shape, such that it} has a large gradient at {mid-radius}. This {renders} the ITG modes unstable in this region of the plasma.
The electron temperature, which is the same as the ion temperature at the maximum gradient position, is $T_e=550$~eV. In this kind of simulation, the energy of the unstable modes increases with time, as no saturation mechanism is included in the simulation. The potential shows typical structures with resonant mode numbers ($m,n$) such that $n/m \approx \iota/2\pi$ and centered around $k_\theta \rho_i \approx \frac{m}{\langle r \rangle} \rho_i \approx 0.5$, $\langle r \rangle$ being the average minor radius of the flux surface at half radius. In this case, $15<m<30$ and $15<n<55$. The unstable modes propagate in the ion diamagnetic direction as corresponds to an ion drift wave.
In this simulation, no zonal flow is generated by these modes, as it is linear and no mode interaction is taken into account. 
Raw simulation data of potential at the locations of the synthetic probes are shown in Fig.~\ref{Euterpe_ITG_data}.

\begin{figure} \centering
  \includegraphics[trim=0 0 0 0,clip=,width=10cm]{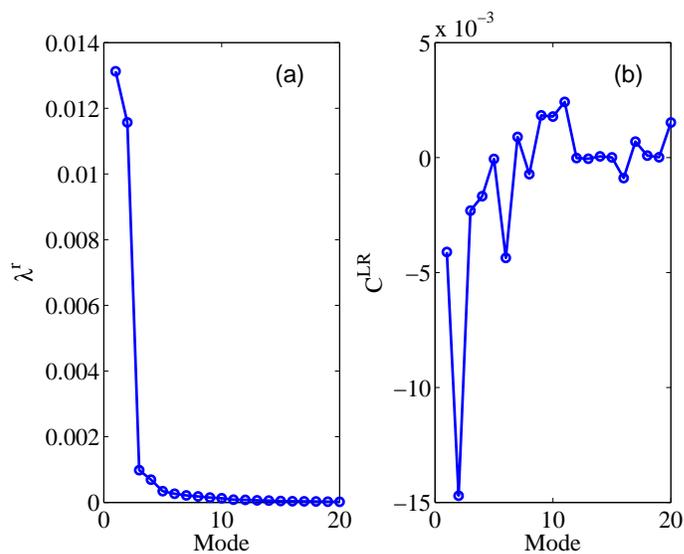}
\caption{\label{Euterpe_ITG_LRC}(a) BOD eigenvalues and (b) the long range correlation contribution of each mode, $C^{\rm LR}$, for an Euterpe simulation of ITG turbulence.}
\end{figure}

The BOD eigenvalues and the long range correlation contribution of each mode, $C^{\rm LR}$, are shown in Fig.~\ref{Euterpe_ITG_LRC}.
By contrast with the preceding section, the LRC is very small here. On the other hand, the first two BOD eigenvalues have a similar amplitude, suggesting a possible mode pair (corresponding to a propagating mode). 

\begin{figure} \centering
  \includegraphics[trim=0 0 0 0,clip=,width=10cm]{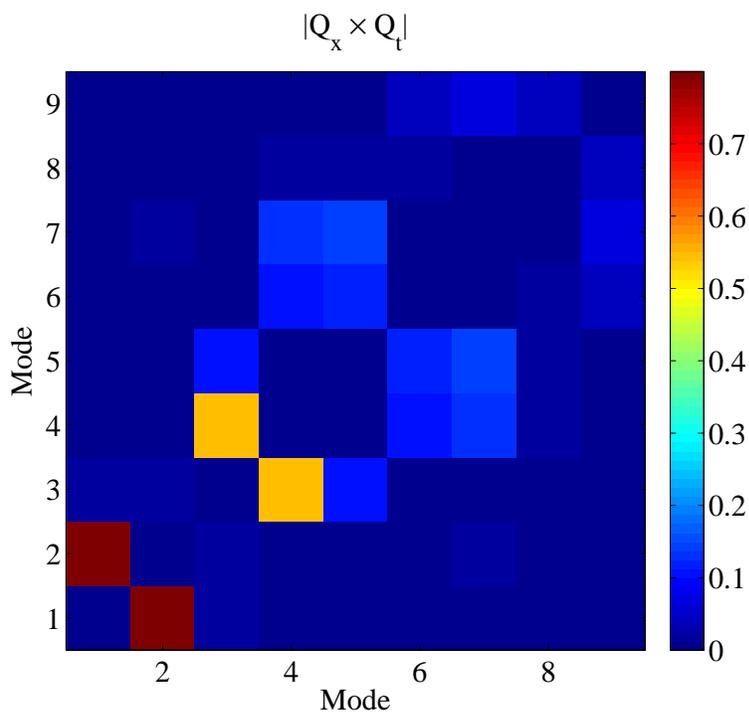}
\caption{\label{Euterpe_ITG_QxQt}Quadrature matrix of the first few BOD modes, for an Euterpe simulation of ITG turbulence.}
\end{figure}

Indeed, the quadrature analysis shown in Fig.~\ref{Euterpe_ITG_QxQt} confirms this suspicion: clearly, modes 1 and 2 form a (propagating) pair, as do modes 3 and 4.

\begin{figure} \centering
  \includegraphics[trim=0 0 0 0,clip=,width=16cm]{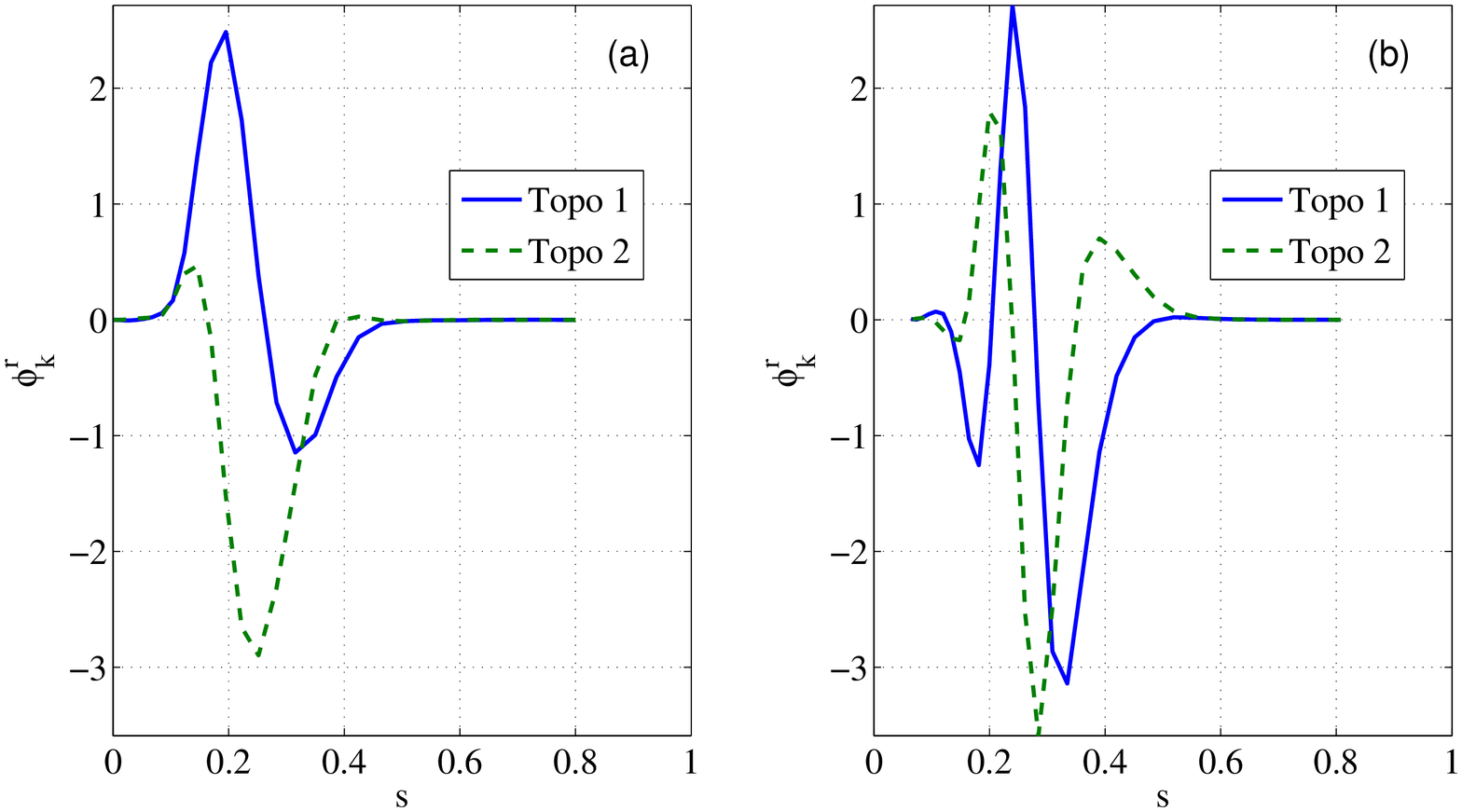}\\
  \includegraphics[trim=0 0 0 0,clip=,width=16cm]{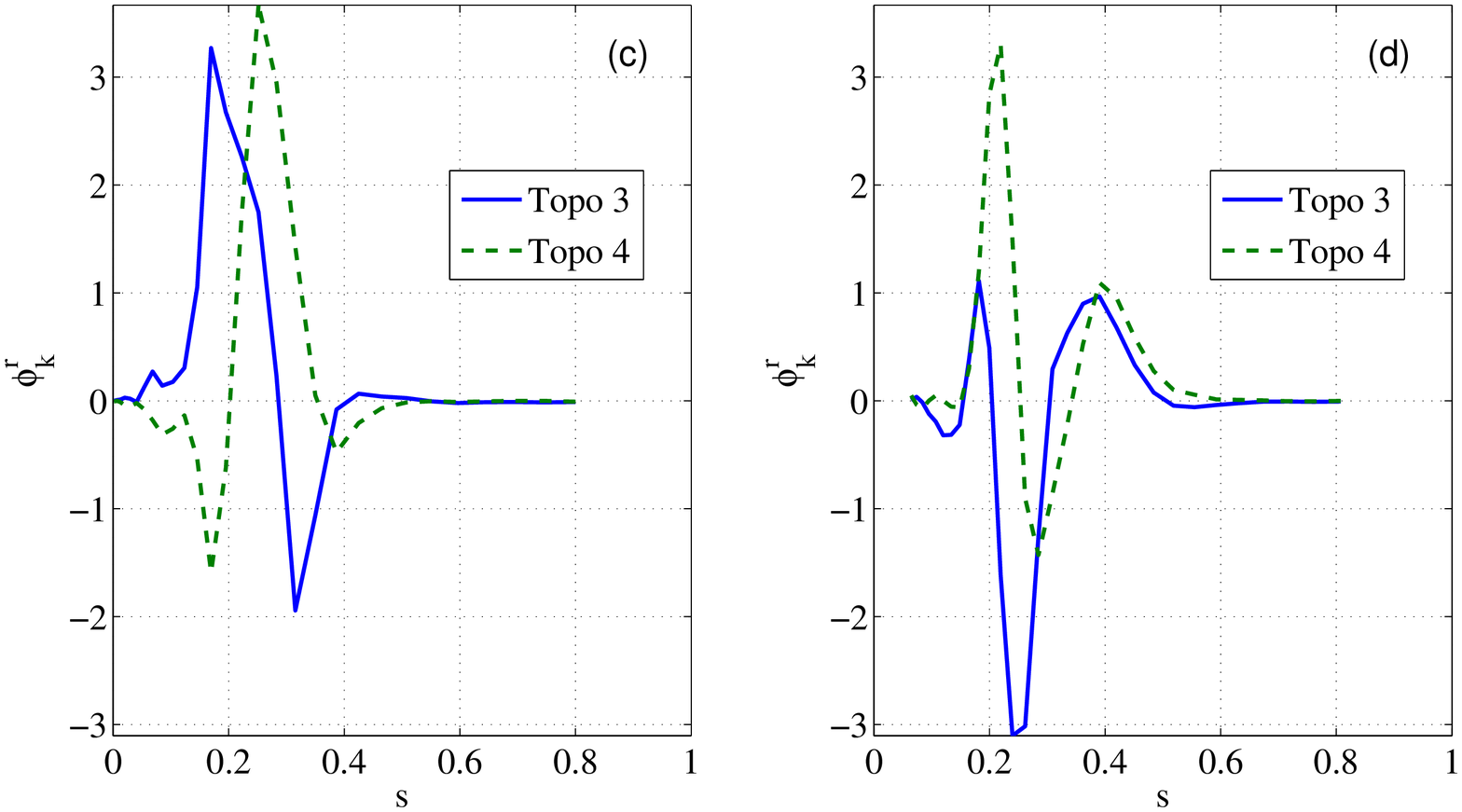}
\caption{\label{Euterpe_ITG_topos}Topos 1 and 2 for (a) probe 1 and (b) probe 2, and topos 3 and 4 for (c) probe 1 and (d) probe 2, for an Euterpe simulation of ITG turbulence.}
\end{figure}

\begin{figure} \centering
  \includegraphics[trim=0 0 0 0,clip=,width=10cm]{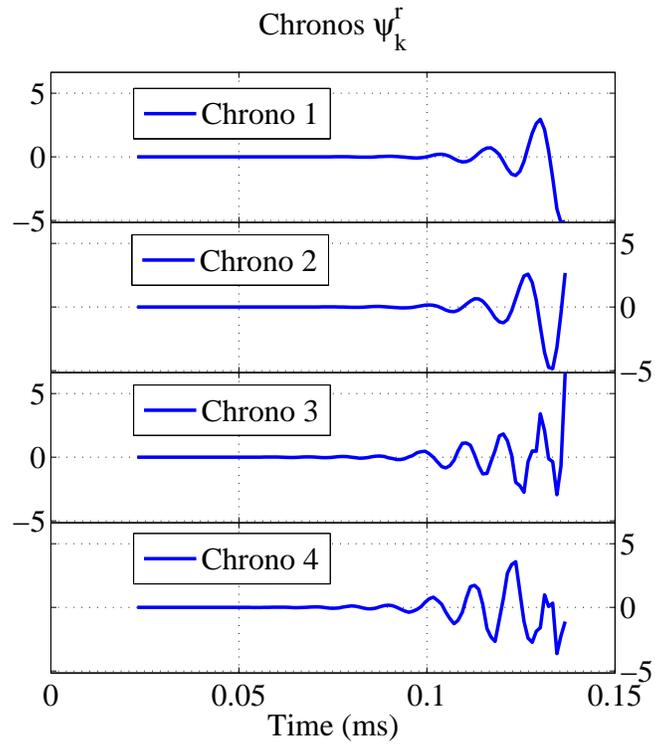}
\caption{\label{Euterpe_ITG_chronos}Chronos, for an Euterpe simulation of ITG turbulence.}
\end{figure}

Fig.~\ref{Euterpe_ITG_topos} shows the topos, reflecting the radial mode structure of these propagating modes.
Fig.~\ref{Euterpe_ITG_chronos} shows the corresponding chronos, which exhibit the expected exponential growth of the oscillation amplitude.
These observations are in accordance with the theoretical mode structure obtained from the distribution of the mode shapes and amplitudes in the simulation.

\clearpage
\section{Application to TJ-II}\label{TJ-II}

In this section, we will apply the techniques described in Section \ref{Method} and tested on gyro-kinetic simulations in Section \ref{Euterpe} to data obtained from a toroidally confined plasma in TJ-II. The goal is to see whether it is actually possible to identify a Zonal Flow with some degree of confidence, using appropriately located potential measurements.

TJ-II disposes of a double reciprocating probe system, allowing the simultaneous measurement of the floating potential on various radially separated probe pins on two probes with a large toroidal separation~\cite{Pedrosa:1999}.
Fluctuating structures with a high level of correlation between the toroidally separated probes (`Long Range Correlation' or LRC) can possibly be identified with Zonal Flow (ZF) structures~\cite{Alonso:2012}. 
The radially spaced pins on each of the two probes provide information about the radial wave number of the detected structures.

However, Zonal Flows are not the only phenomena that can give rise to LRCs.
In particular, it has been shown that other modes (e.g., drift waves or MagnetoHydroDynamic (MHD) oscillations) may also produce LRCs~\cite{Milligen:2013b}.
The question, therefore, is whether one can distinguish between different origins of observed LRCs: on the one hand, ZF-like behavior and on the other, MHD or other (rotating/propagating) oscillations.
ZFs are electrostatic potential fluctuations having toroidal mode number $n=0$, poloidal wave number $m=0$, and a finite radial wavenumber~\cite{Diamond:2005}. Due to this symmetry, the ZF potential structure may fluctuate -- relatively slowly -- but {\it does not rotate}.
The latter property allows one to distinguish a ZF from, e.g., a low-frequency MHD mode, which (with very rare exceptions) usually does rotate.

The two techniques discussed in Section \ref{Method} (i.e., the quantification of the contribution of each mode to LRCs and the quadrature detection technique) should enable one to make this distinction based on the experimental data: ZFs are long-range correlated `standing wave structures', while MHD, rotating or propagating modes are `traveling wave structures'. 

TJ-II is a Heliac type stellarator with four field periods, having major radius $R = 1.5$ m, minor radius $a < 0.2$ m, and toroidal magnetic field $B_0 < 1$ T. 
{The plasmas considered here} are heated by the electron cyclotron resonant heating (ECRH) system, consisting of two beam lines with an injected power of up to 400 kW each.
Electron density is controlled by means of a gas puffing system and the line average electron density typically reaches values of about $\overline{n_e} \simeq 0.6 \cdot 10^{19}$ m$^{-3}$. The central electron temperature is typically $T_e(0) \simeq 300-800$ eV, and the central ion temperature {is} around $T_i(0) \simeq 120 - 150$ eV.

Langmuir probe D is located in a top port at toroidal position $\phi \simeq 35^\circ$, while probe B is located in a bottom port at $\phi \simeq 195^\circ$. Thus, the two probe systems are remote, separated toroidally by a distance of $\sim 5$ m (about half a toroidal turn)~\cite{Pedrosa:2010}. In the experiments analyzed here, each probe is fitted with a `rake' probe head measuring floating potential at $\sim 10$ radially spaced pins. 
Probe pins are separated radially by 3 mm in the D probe, and by 1.7 mm in the B probe;
{probe D covers a radial range of about 3.5 cm, while probe B covers about 1.5 cm.}

\begin{figure} \centering
  \includegraphics[trim=0 0 0 0,clip=,width=10cm]{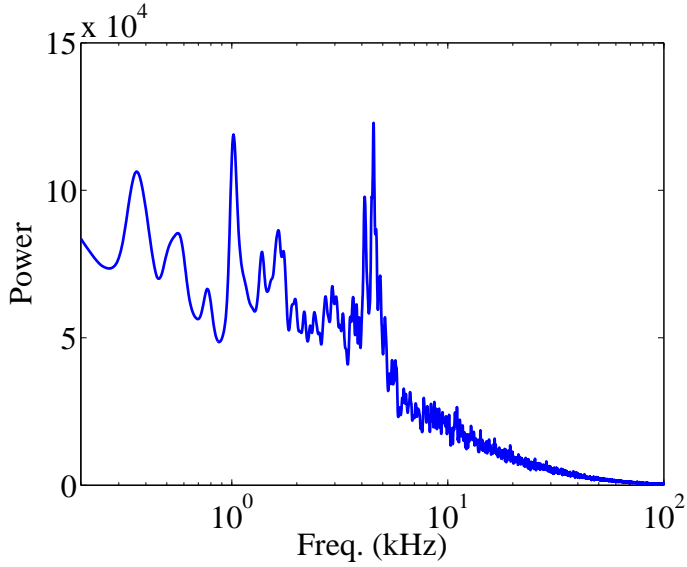}
\caption{\label{36012_spectrum}{Mean spectrum of the probe pins in discharge 36012.}}
\end{figure}

We analyze shot 36012 in the time interval $1090 \le t \le 1250$ ms.
{ In this discharge, the rotational transform, $\iota/2\pi$, has a value of around 1.455 at the magnetic axis and 1.55 at the edge, so that the 3/2 rational surface is located at about $\rho \simeq 0.73$.
In the mentioned} 
time interval, both the line average electron density and the plasma energy content are fairly constant; $\overline{n_e} \simeq 0.63 \cdot 10^{19}$ m$^{-3}$, near the critical density of the electron to ion root confinement transition at TJ-II~\cite{Hidalgo:2006b,Velasco:2012}. 
{Fig.~\ref{36012_spectrum} shows the mean spectrum of all probe pins. Several spectral peaks are visible, which we will identify in the following.}

\begin{figure} \centering
  \includegraphics[trim=0 0 0 0,clip=,width=14cm]{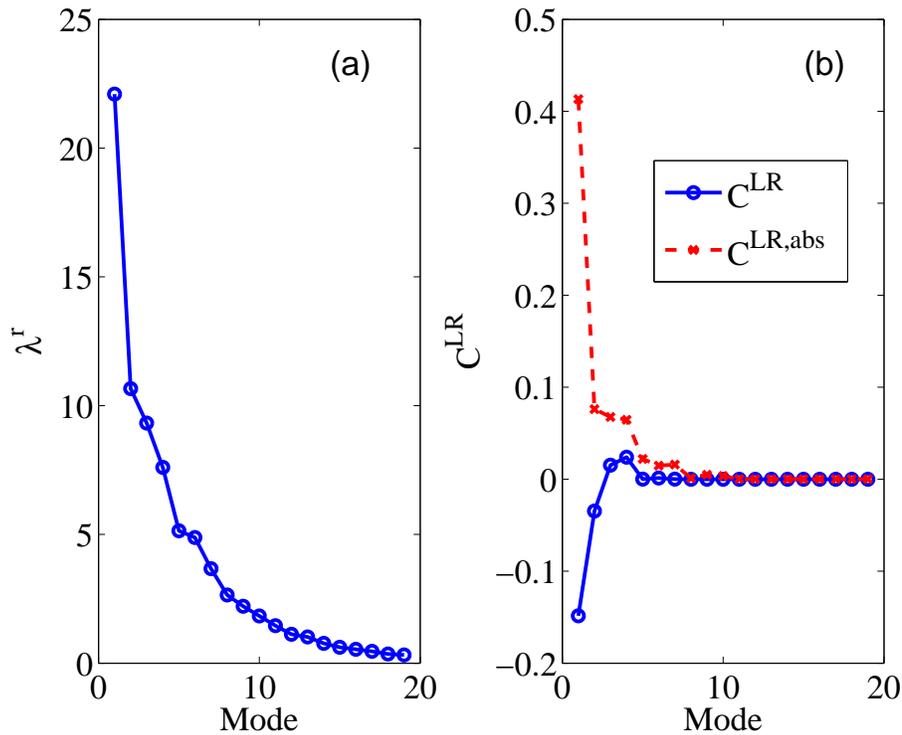}
\caption{\label{36012_lambda}(a) BOD eigenvalues $\lambda_k^r$ and (b) LRC contributions $C_k^{\rm LR}$ and $C_k^{\rm LR, abs}$ (shot 36012).}
\end{figure}

\begin{figure} \centering
  \includegraphics[trim=0 0 0 0,clip=,width=12cm]{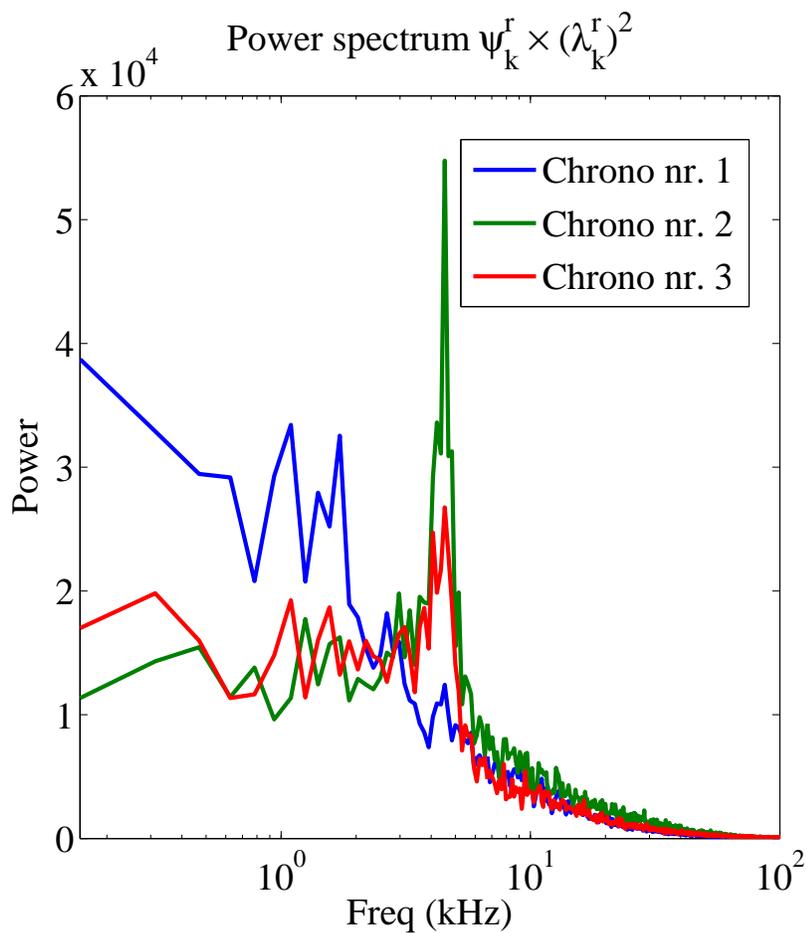}
\caption{\label{36012_spectra}Spectra of the chronos $\psi_k^r$ of the first 3 BOD modes (shot 36012). }
\end{figure}

\begin{figure} \centering
  \includegraphics[trim=0 0 0 0,clip=,width=16cm]{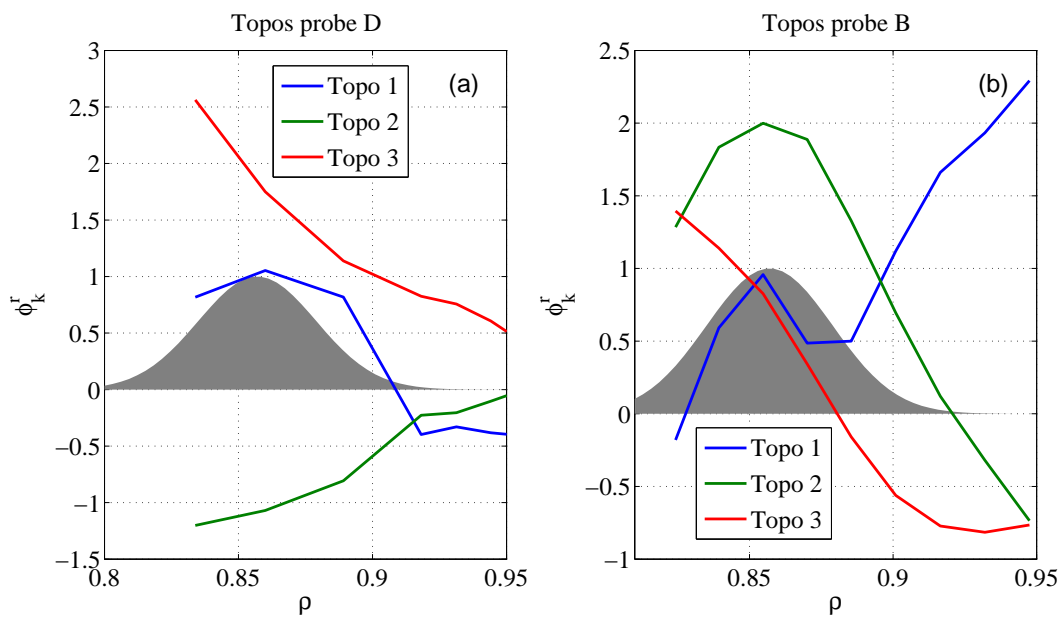}
\caption{\label{36012_topos}Topos $\phi_k^r$ of the first 3 BOD modes for (a) probe D and (b) probe B (shot 36012).}
\end{figure}

Fig.~\ref{36012_lambda} shows the BOD eigenvalues and the LRC.
To calculate the LRC, Eqs.~(\ref{LRC}) and (\ref{LRCabs}), the probe pins have been subdivided into two sets: namely, pins corresponding to probe D or B, respectively.
Fig.~\ref{36012_spectra} shows the spectra of the first 3 modes.
The first BOD mode suggests a strong long range {\it anti}-correlation.
The second and third BOD modes form a propagating pair (confirmed by quadrature analysis, not shown); the spectrum shows that this propagating mode has a clear frequency peak at $5-6$ kHz.
Regarding the negative value of $C^{\rm LR}_1$, it should be noted that topo 1 exhibits a complex radial structure, cf.~Fig.~\ref{36012_topos}. 
Thus, the global character of the definition of $C^{\rm LR}$ may not capture the details of this structure.
The two peaks appearing in topo 1 at $\rho \simeq 0.86$ are in fact {\it correlated} and correspond to a zonal flow-like structure, 
indicated schematically by a grey area in Fig.~\ref{36012_topos}.
This structure is considered zonal flow-like {for} the following reasons:
(1) it is Long Range Correlated (recall Eq.~(\ref{covariance_topos}));
(2) the structure has the same value of floating potential in the two remote probes, suggesting an $m=0, n=0$ global structure:
(3) the spectral characteristics of the structure show no clear peak and spectral power is concentrated at very low frequency, cf.~Fig.~\ref{36012_spectra};
(4) the radial position and radial width are very similar for the two remote probe systems (as indicated by the grey area in Fig.~\ref{36012_topos}).
The radial width of the ZF-like structure is rather small, namely about one centimeter.
The four points mentioned above constitute the most unambiguous identification of a ZF-like structure reported yet in literature.

Further outward ({ $\rho \gtrsim 0.9$}), the floating potentials in the two probes are predominantly  {\it anti-} correlated such that topo 1 has an opposite sign for the two probes.
The temporal variation of this anti-correlated spatial structure is linked to the ZF fluctuations, as is evident from its inclusion in the same chrono.
We speculate that the fluctuations of the ZF intensity (at $\rho \simeq 0.86$) cause a variation of outward transport (in the range { $\rho \gtrsim 0.9$}) with some toroidal/poloidal asymmetry -- probably associated with the different local curvature and flux expansion at the two probe locations.

At this point, we would like to emphasize that the radial structure revealed by this technique is very hard to extract using traditional methods based on, e.g., two-point correlation functions, due to the fact that the signals contain contributions from both the ZF-like structure and the propagating modes. Traditional techniques cannot separate these contributions clearly, unless a hypothesis is made about the spectral distribution and frequency filters are applied. Here, no hypotheses are needed and the results follow directly from the multipoint analysis of the raw data.


\clearpage
\section{Discussion and conclusions}

In this work, we propose a novel analysis method to identify `global', long range correlated (LRC) components from multipoint fluctuation measurements.
The analysis is based on the well-known Biorthogonal Decomposition (BOD), which is extended by defining a quantity that measures the contribution of each BOD mode to the overall LRC, cf.~Eqs.~(\ref{LRC}-\ref{LRCabs}).
In addition, a quadrature detection technique is introduced, based on the Hilbert transform, to determine whether two BOD modes (with similar eigenvalues) are in quadrature and thus are likely to correspond to a propagating mode.
The combination of these various items of information (mode contribution to LRC and quadrature among modes, as well as the spectral and spatial characteristics of the BOD modes) allows distinguishing Zonal Flow-like (`global'), low frequency oscillations from propagating oscillations -- provided sufficient and adequately placed multipoint data are available. 

The method was tested on gyrokinetic simulations with the global code EUTERPE, using data from synthetic probes. 
Two simulations were analyzed, using synthetic probes to generate signals.
In the first simulation, the linear, collisionless relaxation of an initial zonal perturbation to the density was studied.
The BOD analysis successfully extracted the ZF mode from the generated signals.
The GAM-like oscillations appearing at the beginning of the simulation were identified as associated to a ``{radially} propagating structure''.
In the second simulation, the linear ITG instability was simulated, and no ZF was produced.
In this case, the BOD analysis technique detected propagating modes and yielded their structure, and did not detect any ZF (as indeed it shouldn't).

Then, the method was applied to TJ-II Langmuir probe data.
In a discharge near the electron-ion root confinement transition, the method was shown to be able to separate ZF-like fluctuations from propagating contributions to the LRC. 
Unambiguous identification of the ZF-like structure was possible, using four elements:
(1) the structure exhibits (positive) Long Range Correlation;
(2) it has the same value of floating potential in the two remote probes, suggesting an $m=0, n=0$ global structure;
(3) the spectrum shows no clear peak and spectral power is concentrated at very low frequency;
(4) the radial position and radial width are very similar in the two remote probes. 
By contrast, the propagating structure was recognized by the quadrature of two topos and chronos having similar BOD eigenvalues, and a peak in the corresponding spectra.

It is noted that with only two toroidally/poloidally separated multi-pin probes, as used in this work, the capacity to identify modes propagating in the various directions is somewhat limited. 
More sophisticated probes, providing more detailed spatial information in the toroidal/poloidal directions, or the combination of information from various different types of diagnostics might overcome such limitations and allow a better identification of propagating modes.

While bearing this in mind, it is shown that the BOD is a powerful technique to separate ZF-like, global oscillations from other (propagating) oscillations and to extract their mode structure.
In this framework, it has been speculated that microscale turbulence might not only interact with Zonal Flows, but also with MHD modes \cite{Ishizawa:2007,Ishizawa:2008}. This would imply the existence of an (indirect) interaction between ZFs and MHD modes. 
Indeed, in the past it has been observed that the growth of ZFs may modify MHD activity (e.g., \cite{Toi:1989}). 
It is expected that the methodology presented here could help clarifying this issue, or more generally, yield fruitful results for a wide range of multipoint (and remote) measurements in systems with long range correlated structures.

\section*{Acknowledgements}
{The authors thankfully acknowledge the computer resources, technical expertise and assistance provided by the Barcelona Supercomputing Center -- Centro Nacional de Supercomputaci\'on and the CIEMAT Computing Center. }
Research sponsored in part by the Ministerio de Econom\'ia y Competitividad of Spain under project Nr.~ENE2012-30832.
This project has received funding from the European Union's Horizon 2020 research and innovation programme under grant agreement number 633053. 
The views and opinions expressed herein do not necessarily reflect those of the European Commission.

\clearpage 
\section*{References}

\end{document}